\documentclass[12pt,pre,epsfig,address,onecolumn]{revtex4}%
\usepackage[english]{babel}
\usepackage[latin1]{inputenc}
\usepackage[dvips]{graphicx}
\usepackage{amsmath}
\usepackage{amssymb}
\usepackage{epsfig}
\usepackage{array}
\usepackage{amsfonts}
\usepackage{graphicx}
\usepackage{subfigure}
\usepackage{color}%
\newcommand{\be}{\begin{equation}}
\newcommand{\ee}{\end{equation}}
\newcommand{\bel}[1]{\begin{equation}\label{#1}}
\newcommand{\bea}{\begin{eqnarray}}
\newcommand{\eea}{\end{eqnarray}}
\newcommand{\ba}{\begin{array}}
\newcommand{\ea}{\end{array}}

\begin{document}
\title{Transition probabilities and dynamic structure factor in the ASEP conditioned
on strong flux}
\author{V. Popkov$^{1,3}$ and G. M. Sch\"utz$^{2,3}$ }
\affiliation{$^{1}$ Dipartimento di Fisica "E.R. Caianiello", and
Consorzio Nazionale Interuniversitario per le Scienze Fisiche
della Materia (CNISM), Universit\`a di Salerno, Fisciano, Italy}
\affiliation{$^{2}$ Institut f\"ur Festk\"orperforschung,
Forschungszentrum J\"ulich, 52425 J\"ulich, Germany}
\affiliation{$^{3}$Interdisziplin\"ares Zentrum f\"ur Komplexe
Systeme, Universit\"at Bonn, Br\"uhler Stra{\ss}e 7, 53119 Bonn,
Germany}
\date{\today }

\begin{abstract}
We consider the asymmetric simple exclusion processes (ASEP) on a ring
constrained to produce an atypically large flux, or an extreme activity. Using
quantum free fermion techniques we find the time-dependent conditional
transition probabilities and the exact dynamical structure factor under such
conditioned dynamics. In the thermodynamic limit we obtain the explicit
scaling form. This gives a direct proof that the dynamical exponent in the
extreme current regime is $z=1$ rather than the KPZ exponent $z=3/2$ which
characterizes the ASEP in the regime of typical currents. Some of our results
extend to the activity in the partially asymmetric simple exclusion process,
including the symmetric case.

\end{abstract}
\maketitle


\section{Introduction}

\label{sec::Introduction}

The Asymmetric Simple Exclusion Process (ASEP) is a minimal model for many
traffic and queueing processes \cite{Ligg99,Schu01,DerridaReview2007}. It
describes a non-equilibrium system of many driven particles, interacting via
hard core repulsion. This Markov process is defined on a lattice, each site of
which can be empty or occupied by one particle. Particles jump independently
after an exponentially distributed random time with mean $1/(p+q)$ to a nearest
neighbor site on the right or on the left, provided that the target site is
empty (hard core exclusion rule). The probability to choose the right
neighbour is $p/(p+q)$ while the probability of choosing the left neighbour is
$q/(p+q)$. In spite of its simple formulation, and very simple product
stationary state, the time-dependent characteristics are very nontrivial to
obtain, even in the simplest periodic one-dimensional case (closed ring of $L$
sites). This is due to the fact that e.g. the time-dependent operators of type
$\mathrm{e}^{Ht}\hat{f}(0)\mathrm{e}^{-Ht}$ are complicated objects because
the generator of the stochastic dynamics $H$ is equivalent to an interacting
many-body quantum Hamiltonian.

Recently, it has been found that the ASEP, under the restriction
to produce an extreme flux or extreme activity is governed by
effective long range interactions \cite{ExtremeASEP_JStat2010}. In
this setting one considers realizations of the process for a
duration $T$ which for a long interval between  large times $t$
and $T-t$ (where $T-t$ is itself also large) have carried an
atypically large flux. This extreme event quite surprisingly makes
the conditioned process intrinsically related to a much simpler
system than the original one, namely to a system of
non-interacting fermions. This fact suggests to use field
theoretic free fermion techniques to compute time-dependent
correlations in this particular large deviation limit of the
classical stochastic dynamics of the conditioned ASEP.

In the present contribution we shall employ these techniques to calculate two
particular time-dependent correlation functions: the conditional probabilities
$P(\eta,t;\eta_{0},0)$ for the transition from a microscopic $N$-particle
initial configuration $\eta_{0}$ to a final configuration $\eta$ at time $t$
and the dynamical structure factor which is the Fourier transform of the
time-dependent density-density correlation function $\langle n_{k}%
(t)n_{0}(0)\rangle-\rho^{2}$ in the stationary ASEP in the two above mentioned
limits: under the restriction to produce a very large flux and under the
restriction to produce a very large activity. We note that for the totally
asymmetric version of the ASEP (TASEP) on an infinite lattice and on a ring
the conditional probabilities were found in \cite{Schu97InfiniteTASEP} and
\cite{PriezzhevRingTASEP} respectively using the Bethe ansatz. The dynamical
structure for the infinite system was obtained by Pr\"ahofer and Spohn using
random matrix techniques \cite{Prae02}. We remark that that the time-dependent
correlations for atypical stochastic trajectories (in the large deviation limit)
are practically unaccessible by numerical methods, even though some promising
approaches have been developed recently \cite{GiardinaKurchanPeliti2006}.

The paper is organized as follows. In Sec. \ref{condition} we introduce the
ASEP conditioned to carry a certain nontypical flux or to exhibit a nontypical
activity. In the first two subsections we review for self-containedness the
our previous work \cite{ExtremeASEP_JStat2010} which among other things
clarifies how conditioning on large flux and activity is related to free
fermions. In the third subsection of Sec. \ref{condition} we present a new
result, viz. the form of temporal correlations in the large current or large
activity regime. In the following main sections we focus on large flux and
derive the conditional transition probabilities (Sec. \ref{probabilities}) and
dynamical structure factor (Sec. \ref{structure}). We conclude with a brief
summary and some remarks on the large limit activity in the ASEP and symmetric exclusion.

\section{ASEP conditioned on a large flux}

\label{condition}

\subsection{Definitions and relation to an exclusion process with long-range interaction}

The flux (or synonymously the current) in the ASEP is the mean net
number of jumps in some time interval. Conditioning the ASEP to
carry an typical flux is achieved by first ascribing to the ASEP
space-time trajectories weighting factors $\mathrm{e}^{sJ}$ where
$s$ is a generalized chemical potential, conjugated to the total
current $J$, registered along the trajectory up to time $t$ (see
e.g. \cite{DerridaAppert99} for a elaborate treatment). Fixing
$s$, one picks up an ensemble of space-time trajectories
characterized by any desired average current $j(s)$. In
particular, fixing the desired current to be very large  (in the
sense described in the introduction), one obtains a particular set
of trajectories, where several tendencies are clearly seen
\cite{ExtremeASEP_JStat2010} when taking the limit
$s\rightarrow\infty$: (a) the backward hoppings are totally
suppressed (b) formation of clusters of particles, potentially
leading to the flux reduction, is strongly suppressed (c) long
range interactions between the particles are generated.

In more details, it was noted that the ASEP under the restriction to produce
extreme flux after proper rescaling of time is described by a Master equation
\begin{equation}
\frac{\partial|P\rangle}{\partial t}=-H_{eff}|P\rangle\label{MasterEquation}%
\end{equation}
with effective stochastic Hamiltonian%
\begin{equation}
H_{eff}=\Delta H\Delta^{-1}-\mu, \label{Heff}%
\end{equation}
where $\Delta$ is a diagonal matrix with positive entries given further below,
and $\mu$ is the lowest eigenvalue of the Hamiltonian $H$
\begin{equation}
H=-\sum_{n=1}^{L} \sigma_{n}^{+}\sigma_{n+1}^{-} . \label{UniDirXX0}%
\end{equation}
with the spin 1/2 lowering and raising operators of the Lie algebra $SU(2)$.
We point out that in the limit of very large flux the same stochastic process
describes the partially asymmetric simple exclusion process (ASEP) or even the
symmetric case, since in this limit any backward hopping does not contribute.

Similar considerations can be carried out for the activity where space-time
trajectories are weighted by a factor $\mathrm{e}^{sA}$ where $A$ is the total
number of jumps along the trajectory, irrespective of the direction of the
jump. In the case of ASEP restricted to have an extremely large activity
(counting hoppings, irrespective of the direction), the effective Hamiltonian
has the form (\ref{Heff}) where $H$ is substituted by
\begin{equation}
H_{II}=-\sum_{n=1}^{L}\left(  p\sigma_{n}^{+}\sigma_{n+1}^{-}+q\sigma
_{n+1}^{+}\sigma_{n}^{-}\right)  , \label{ASEPactivity}%
\end{equation}
and $\mu$ is \ substituted by the lowest eigenvalue of $H_{II}$ and $p$ is the
hopping rate to the right (clockwise) while $q$ is the hopping rate to the
left (anti-clockwise) of the ASEP. Both Hamiltonians (\ref{UniDirXX0}%
),(\ref{ASEPactivity}) are free fermion Hamiltonians, and in particular for
the case of symmetric hopping $p=q=1$, the so called Symmetric Simple
Exclusion Process, or SSEP, the dynamics is governed by the effective
Hamiltonian (\ref{Heff}) with $H$ \ substituted by the physical $XX0$
Hamiltonian describing quantum fermions on a ring of $L$ sites,
\begin{equation}
H_{XX0}=-\sum_{n=1}^{L}\left(  \sigma_{n}^{+}\sigma_{n+1}^{-}+\sigma_{n+1}%
^{+}\sigma_{n}^{-}\right)  \label{XX0}%
\end{equation}
and $\mu$ \ substituted by the lowest eigenvalue of $H_{XX0}$. All the above
mentioned Hamiltonians commute with each other and share the same ground state
$|\mu\rangle$. The components of $|\mu\rangle$ in the natural basis constitute
the respective entries of the diagonal matrix $\Delta$ from (\ref{Heff})
$\langle\eta|\mu\rangle=\Delta_{\eta\eta}$. By the Perron-Frobenius theorem
the ground state is nondegenerate, all its components are positive
$\langle\eta|\mu\rangle$, so the matrix $\Delta^{-1}$ exists. It is
straightforward to verify that the stationary state (the state with zero
eigenvalue) of the stochastic process (\ref{MasterEquation}) is given by
$\Delta|\mu\rangle$. The stationary state is thus common for all Hamiltonians
mentioned above, for details see \cite{ExtremeASEP_JStat2010},\cite{Simo09}.

 It may be worthwhile noting that the effective process is an interesting
object of study in its own right, i.e., without reference to atypically large
currents in the usual ASEP. This process is a totally asymmetric exclusion
process with long-range interaction. It describes nearest neighbour hopping,
but the hopping rate depends on the positions of the other particles.
Specifically, the move of the $k$-th particle, located at position $n_{k}$ in
a configuration $\eta$, to the consecutive position $n_{k}+1$ in the new
configuration $\eta^{\prime}$ has the rate \cite{ExtremeASEP_JStat2010}
\begin{equation}
\label{hoppingrates}
W_{\eta^{\prime}\eta}=\prod\limits_{l\neq k} \frac{\sin(\pi(n_{k}+1-n_{l})/L)}
{\sin(\pi(n_{k}-n_{l})/L)}%
\end{equation}
where the product is over $l$ and $k$ from 1 through $N$. If the target site
$n_{k}+1$ is occupied, the hopping rate is zero, in agreement with the
exclusion rule.

For the case of extreme activity one obtains in the same fashion a
partially asymmetric or symmetric exclusion process with long
range interaction. The symmetric case has been studied some time
ago by Spohn \cite{Spoh99} who derived a hydrodynamic fluctuation
theory for the large scale dynamics of the process. Noticing that
fluctuations are driven by a Gaussian field he presented a general
form of the dynamical structure function which represents a
universality different from the Edwards Wilkinson equation. He
also noted the link of the hopping rates (\ref{hoppingrates}) to
Dyson's Brownian motion, which was pointed out independently in
\cite{ExtremeASEP_JStat2010} for the totally asymmetric case. Thus
our discrete model has a natural interpretation as a discrete,
biased random walk version of Dyson's Brownian motion driven by an
external field.

\subsection{Free Fermion states and spectrum}

The complete set of eigenstates of $H,H_{II}$ and $H_{XX0}$ in a
sector with $N$ particles is characterized by a combination of $N$
plane waves where each plane has a quasimomentum $\alpha$
satisfying $\mathrm{e}^{i\alpha L}=(-1)^{N+1}$ because of the
periodic boundary conditions \cite{Gaud81}. Therefore each
$\alpha$ takes one of the quantized values
\begin{equation}
\alpha_{k}=\frac{2\pi}{L}(k-\frac{N+1}{2}),\quad k\in\{1,\dots,L\}
\label{quasimomenta}%
\end{equation}
and thus each eigenstate is defined by an $N$-tuple of integers $\{k\}\equiv
\{k_{1},\dots,k_{N}\}$ where each $k_{j}\in\{1,\dots,L\}$. With these
definitions we can write the eigenvectors as
\begin{equation}
|\mu_{\{k\}}\rangle=\sum_{m_{1}m_{2}...m_{N}}\chi_{m_{1}m_{2}...m_{N}%
}(\{k\})|m_{1}m_{2}....m_{N}\rangle\label{mu_q}%
\end{equation}
where the vector $|m_{1}m_{2}....m_{N}\rangle$ denotes a state with particles
at respective positions $m_{j}$ (not necessarily ordered) and the sum is over
all possible particle positions $m_{j}\in\{1,2,...,L\}$. The free fermion wave
function in coordinate representation has the form
\begin{equation}
\chi_{m_{1}m_{2}...m_{N}}(\{k\})=\frac{1}{N!L^{N/2}}T_{\{m\}}\sum_{Q}%
(-1)^{Q}\mathrm{e}^{i\sum_{j=1}^{N}m_{Q_{j}}\alpha_{k_{j}}} \label{ksi}%
\end{equation}
where $Q$ is a permutation of indexes $Q(1,2,...,N)=(Q_{1},Q_{2},...,Q_{N})$
and $(-1)^{Q}$ denotes the sign of the permutation. The factor $T_{\{m\}}$ is
zero if in the set $\{m_{1}m_{2}....m_{N}\}$ some $m_{j}$ coincide, and
otherwise is equal to $1$ or $-1$, depending on whether the set $\{m_{1}%
m_{2}....m_{N}\}$ was obtained from the ordered set by an even or odd number
of pair permutations. Any exchange of particle positions or quasimomenta
results in the eigenfunction $|\mu_{\{k\}}\rangle$ changing sign, which
reflects the Pauli principle. The states $|\mu_{\{k\}}\rangle$ with all
possible sets $\{k\}$ are normalized, orthogonal and form an orthonormal basis
in the respective sector of Hilbert space with $N$ number of particles. The
eigenvalues $E^{\{k\}}$ corresponding to the $|\mu_{\{k\}}\rangle$ are sums of
those for each quasiparticle
\begin{equation}
E_{\{k\}}^{Z}=\sum_{j=1}^{N}\varepsilon^{Z}(\alpha_{k_{j}}) \label{E}%
\end{equation}
where the quasiparticle energies are
\begin{align}
\varepsilon^{I}(\alpha_{k})  &  =-\mathrm{e}^{-i\alpha_{k}}%
\label{defaultenergy}\\
\varepsilon^{II}(\alpha_{k})  &  =-(p\mathrm{e}^{-i\alpha_{k}}+r\mathrm{e}%
^{i\alpha_{k}})\label{E_II}\\
\varepsilon^{XX0}(\alpha_{k})  &  =-2\cos\alpha_{k} \label{E_XX0}%
\end{align}
for the Hamiltonians (\ref{UniDirXX0}),(\ref{ASEPactivity}) and (\ref{XX0}), respectively.

The ground state $|\mu_{0}\rangle$ which minimizes the energies $E_{\{k\}}%
^{Z}$ is characterized by the choice of quasimomenta $\alpha_{k}$ where
$k_{j}=j$, i.e. for the set
\begin{equation}
\{k\}_{0}=\{1,2,...N\}. \label{GSmomenta}%
\end{equation}
For this choice the ground state energies $E_{0}^{Z}$ for all three cases are
real. In the following by default we treat the Hamiltonian (\ref{UniDirXX0})
with quasiparticle "energies" (\ref{defaultenergy}) and drop the superscript
$I$.

\subsection{Correlation Functions}

Consider a general time-dependent correlation function $\langle
G(t)F(0)\rangle_{eff}$ in the stationary distribution of the effective
stochastic process defined by (\ref{Heff}). Here $G$ and $F$ are functions of
the occupation numbers and hence represented in the quantum Hamiltonian
formalism by diagonal operators and $G(t) = \exp(H_{eff}t)G\exp(-H_{eff}t)$.
The basis of our present work is the following simple, but fundamental
property of conditioned processes of the form (\ref{Heff}).

\textbf{Theorem}. Let the operators $G,F$ be diagonal in the natural basis
defined by $H_{eff}$. Then
\begin{equation}
\langle G(t)F(0)\rangle_{eff}=\langle\mu| \tilde{G}(t)F| \mu\rangle\label{Th}%
\end{equation}
where $\tilde{G}(t)=\exp(Ht)G\exp(-Ht)$ and $H$ is the respective free fermion
Hamiltonian (\ref{UniDirXX0}),(\ref{ASEPactivity}) or (\ref{XX0}).

Proof. The proof is simple, but for readers not familiar with the
machinery we provide here the details. By definition,
\begin{equation}
\langle G(t)F(0)\rangle_{eff}=\langle s|\mathrm{e}^{H_{eff}t}G\mathrm{e}%
^{-H_{eff}t}F|P^{\ast}\rangle\label{Th2}%
\end{equation}
where the summation vector $\langle s|$ and stationary distribution vector
$|P^{\ast}\rangle$ are left and right lowest eigenvectors of $H_{eff}$ with
eigenvalue 0. We recall that $|P^{\ast}\rangle=\Delta|\mu\rangle$
\cite{ExtremeASEP_JStat2010} and that the summation vector $\langle
s|=\langle1111...1|$ is a vector with all unit components. The property
$\langle s|H_{eff}=0$ is simply a stochasticity condition (conservation of
probability) for the Hamiltonian $H_{eff}$. Using the latter condition and the
fact that $[\Delta,F]=0$ when both $\Delta$ and $F$ diagonal operators we
obtain
\begin{equation}
\langle G(t)F(0)\rangle_{eff}=\langle s|G\mathrm{e}^{-H_{eff}t}\Delta
F|\mu\rangle. \label{Th3}%
\end{equation}
Now we note that $\mathrm{e}^{-H_{eff}t}\Delta=\Delta\mathrm{e}^{-(H-\mu)t}$,
which is verified by expanding the exponent on both sides in Taylor series and
comparing the series term by term, using (\ref{Heff}). Inserting this equality
into (\ref{Th3}), and using $[\Delta,G]=0$, we obtain
\begin{equation}
\langle G(t)F(0)\rangle_{eff}=\langle s|\Delta G\mathrm{e}^{-(H-\mu)t}%
F|\mu\rangle. \label{Th4}%
\end{equation}
Finally, noting $\langle s|\Delta=\langle\mu|$ and $\langle\mu|\mathrm{e}^{\mu
t}=\langle\mu|\mathrm{e}^{Ht}$, we obtain the right hand side of (\ref{Th}).
\hfill$\Box$

Some remarks are in order. Theorem (\ref{Th}) is straighforwardly generalized
to multipoint space-time correlation functions and one obtains $\langle
G(t_{1})F(t_{2})...Q(t_{n})\rangle_{eff}= \langle\mu| \tilde{G}(t_{1})
\tilde{F}(t_{2})\dots\tilde{Q}(t_{n})|\mu\rangle$, provided that all operators
$G,F,...,Q$ are diagonal in the natural basis, i.e., represent observables of
the classical stochastic process generated by $H_{eff}$. We also point out
that the theorem is phrased for our specific case of the current or activity
in the ASEP. However, a similar results applies for any conditioned stochastic
dynamics in which case $H \equiv H(s)$ is the weighted generator of the
unconditioned process.

\section{Conditional Probabilities}

\label{probabilities}

We pick two arbitrary configurations $\eta_{0}$ and $\eta$ of our system
containing $N$ particles at the positions $n_{i}$ and $m_{i}$ respectively,
i.e., $|\eta_{0}\rangle=|n_{1}n_{2}....n_{N}\rangle$ , $|\eta\rangle
=|m_{1}m_{2}....m_{N}\rangle$. We assume both sets to be ordered,
$n_{i}<n_{i+1}$ and $m_{i}<m_{i+1}$. The probability to find the system in
configuration $\eta$ at time $t$, provided it has been in configuration
$\eta_{0}$ at time $t=0$ is given by
\begin{equation}
P_{L}(\eta,t;\eta_{0},0)=\frac{\langle\hat{\eta}(t)\hat{\eta}_{0}\rangle
_{eff}}{\langle\hat{\eta}_{0}\rangle_{eff}}. \label{CondProb}%
\end{equation}
where the subscript denotes that the average is computed with respect to the
stationary state defined by the effective dynamics (\ref{MasterEquation}). The
operators $\hat{\eta}(0)=|\eta\rangle\langle\eta|$ and $\hat{\eta}_{0}%
=|\eta_{0}\rangle\langle\eta_{0}|$ are diagonal operators, which allows to use
the Theorem (\ref{Th}). The denominator in (\ref{CondProb}) can be obtained by
using (\ref{Th}) with $G=I$ and $F=\hat{\eta}_{0}$. Using the results for the
stationary probabilities of ref. \cite{ExtremeASEP_JStat2010} we find
\begin{equation}
\langle\hat{\eta}_{0}\rangle_{eff}=\langle\mu|\eta_{0}\rangle\langle\eta
_{0}|\mu\rangle=(N!)^{2}\chi_{\eta_{0}}^{\ast}\chi_{\eta_{0}}=\frac
{2^{N(N-1)}}{L^{N}}\prod\limits_{\substack{i,j\\1\leq i<j\leq N}}\sin^{2}%
\pi\frac{n_{i}-n_{j}}{L} \label{eta0}%
\end{equation}
The right hand side of (\ref{eta0}) are (modulo squares) the components of a
Slater determinant with quasimomenta (\ref{GSmomenta}) chosen so as to fill
the Fermi sea. For details, see e.g. \cite{KorepinTMP93,Gaud81}.

For the numerator we have, using (\ref{Th}),%
\begin{equation}
\langle\hat{\eta}(t)\hat{\eta}_{0}\rangle_{eff}=\langle\mu|\mathrm{e}^{Ht}%
\eta\mathrm{e}^{-Ht}\eta_{0}|\mu\rangle. \label{Nom}%
\end{equation}
Since the eigenvectors (\ref{mu_q}) form an orthonormal basis the sum of
projectors $(N!)^{-1}\sum_{\{k\}}|\mu_{\{k\}}\rangle\langle\mu_{\{k\}}|=I$
over all $N$-tuples $\{k\}$ is a unit operator. The factor $(N!)^{-1}$ appears
because in the sum $\sum_{\{k\}}f(\{\alpha_{k}\})=\sum_{j_{1}=1}^{L}%
...\sum_{j_{N}=1}^{L}f(\alpha_{j_{1}},\alpha_{j_{2}},...\alpha_{j_{N}})$ each
different set of $\alpha$-s occurs $N!$ times. Inserting it in (\ref{Nom}), we
obtain%
\begin{equation}
\frac{1}{N!}\langle\mu|\mathrm{e}^{Ht}\eta\mathrm{e}^{-Ht}\sum_{\{k\}}%
|\mu_{\{k\}}\rangle\langle\mu_{\{k\}}|\eta_{0}|\mu\rangle=\frac{1}{N!}%
\sum_{\{k\}}\mathrm{e}^{(E_{0}-E_{\{k\}})t}Z^{\ast}(\eta)Z(\eta_{0}),
\label{Nom1}%
\end{equation}
where $E_{0}$ is the ground state energy corresponding to the specific set of
quasimomenta $\{k\}_{0}$ given by (\ref{GSmomenta}), and $Z(\eta)=\langle
\mu_{\{k\}}|\eta|\mu\rangle=(N!)^{2}\chi_{\eta}^{\ast}(\{k\})\chi_{\eta
}(\{k\}_{0})$ are found using (\ref{mu_q}). Substituting $Z(\eta)$ into
(\ref{Nom1}), one obtains
\begin{equation}
\langle\hat{\eta}(t)\hat{\eta}_{0}\rangle_{eff}=(N!)^{3}\chi_{\eta}^{\ast
}(\{k\}_{0})\chi_{\eta_{0}}(\{k\}_{0})\mathrm{e}^{Et}\sum_{\{k\}}%
\mathrm{e}^{-E_{\{k\}}t}\chi_{\eta_{0}}^{\ast}(\{k\})\chi_{\eta}(\{k\})
\label{Nom2}%
\end{equation}
Using the explicit form of $\chi$ (\ref{ksi}), the sum over $\{k\}$ in
(\ref{Nom2}) can be rewritten as
\begin{equation}
\frac{L^{-N}}{(N!)^{2}}\sum_{\{k\}}\sum_{Q,Q^{\prime}}(-1)^{Q+Q^{\prime}%
}\mathrm{e}^{i\sum_{j=1}^{N}(m_{Q_{j}}-n_{Q_{j}^{\prime}})\alpha_{k_{j}}%
-t\sum_{j=1}^{N}\varepsilon(\alpha_{k_{j}})}. \label{Nom3}%
\end{equation}
In the sum over the permutations $\sum_{Q,Q^{\prime}}$ there are $(N!)^{2}$
terms; however, under the summation over $k_{1},k_{2},...k_{N}$ and
reshuffling of $k$-s only $N!$ terms are independent, namely, $\sum
_{\{k\}}\sum_{Q,Q^{\prime}}(-1)^{Q+Q^{\prime}}\mathrm{e}^{i\sum(m_{Q_{j}%
}-n_{Q_{j}^{\prime}})\alpha_{k_{j}}-t\sum\varepsilon(\alpha_{k_{j}})}%
=N!\sum_{\{k\}}\sum_{Q}(-1)^{Q}\mathrm{e}^{i\sum(m_{Q_{j}}-n_{j})\alpha
_{k_{j}}-t\sum\varepsilon(\alpha_{k_{j}})}$. Substituting this into
(\ref{Nom3}), we get
\begin{equation}
\frac{L^{-N}}{N!}\sum_{\{k\}}\sum_{Q}(-1)^{Q}\mathrm{e}^{i\sum_{j=1}%
^{N}(m_{Q_{j}}-n_{j})\alpha_{k_{j}}-t\sum_{j=1}^{N}\varepsilon(\alpha_{k_{j}%
})}. \label{Nom4}%
\end{equation}

So far the discussion has been general and applicable to all three
Hamiltonians (\ref{UniDirXX0}),(\ref{ASEPactivity}) and (\ref{XX0}). Focussing
now on the default case (\ref{UniDirXX0}) we use (\ref{E}) and
(\ref{defaultenergy}) for further simplification. We expand the exponents
$\mathrm{e}^{-t\varepsilon(\alpha_{k})}=\sum_{s=0}^{\infty}t^{s}%
\mathrm{e}^{i\alpha_{k}s}/s!$ in (\ref{Nom4}), and collect the terms with the
same $\alpha_{k}$. Under the summation over $\{k\}$, only the terms with
$m_{Q_{j}}-n_{j}-s=-\kappa L$ contribute to the sum, where $\kappa=0,1,2,...$
if $m_{Q_{j}}\geq n_{j}$ and $\kappa=1,2,..$ otherwise. Since each $\alpha
_{k}$ satisfies $\mathrm{e}^{i\alpha_{k}L}=(-1)^{N+1}$, for odd number of
particles $N$ each contributing term in (\ref{Nom4}) is equal to $\sum
_{\{k\}}1=L^{N}.$ For even $N$, the signs of the terms with different $\kappa$
will alternate. Now we introduce the function
\begin{equation}
g_{L}(d,t)=\sum_{\kappa=0}^{\infty}\left[  (-1)^{\kappa}sign(d)\right]
^{N+1}\frac{t^{d_{L}+\kappa L}}{(d_{L}+\kappa L)!},\label{g(d)}%
\end{equation}
where $d$ is an integer ranging from $-L+1$ to $L-1$ and $d_{L}=d$ for $d>0$
and $d_{L}=d+L$ for $d<0$. The function $sign(d)=1$ for $d\geq0$ and
$sign(d)=-1$ for $d<0$.  With this function the expression (\ref{Nom4}) can be
rewritten in compact determinantal form as
\begin{equation}
\frac{L^{-N}L^{N}}{N!N!}\sum_{Q}(-1)^{Q}\prod\limits_{k=1}^{N}g_{L}(m_{Q_{k}%
}-n_{k},t)=\frac{1}{(N!)^{2}}\det[g_{L}(m_{j}-n_{i},t)]\label{Nom5}%
\end{equation}
where we used Leibnitz formula for the determinant. Finally, using
(\ref{eta0}), (\ref{Nom2}) and (\ref{Nom5}), the conditional probabilities
(\ref{CondProb}) can be brought after some algebra into the final form
\begin{equation}
P_{L}(\eta,t;\eta_{0},0)=\mathrm{e}^{E_{0}t}\sqrt{\frac{\langle\eta
\rangle_{eff}}{\langle\eta_{0}\rangle_{eff}}}\det[g_{L}(m_{j}-n_{i}%
,t)]\label{CondProbRes}%
\end{equation}
The $\langle\eta_{0}\rangle_{eff},\langle\eta\rangle_{eff}$ are
stationary probabilities of the initial and the final state
respectively, given by (\ref{eta0}). At the last step of the
calculation we have used the fact that all the components
$\chi_{\eta}(\{k\}_{0})$ of the ground state eigenvector can be
made positive, see the discussion after Eq. (\ref{XX0}).
Consequently
$\chi_{\eta}^{\ast}((\{k\}_{0})/\chi_{\eta_{0}}^{\ast}((\{k\}_{0})=\chi_{\eta
}((\{k\}_{0})/\chi_{\eta_{0}}((\{k\}_{0})=\sqrt{\langle\eta\rangle
_{eff}/\langle\eta_{0}\rangle_{eff}}$. Eq. (\ref{CondProbRes}) is
the main result of this section.  The determinantal structure of
the conditional probabilities was first noticed by Spohn
\cite{Spoh99} for the symmetric case. In contrast to the symmetric
case, here the matrix elements of the determinant are the
propagators of a totally asymmetric random walk. Extending the
link of the symmetric model to Dyson's Brownian motion we may
regard our model as a totally asymmetric Dyson random walk.

Let us discuss some limiting cases of (\ref{CondProbRes}). For $t=0$ one has
$\det[g_{L}(d_{ij},t)]=\delta_{\eta\eta_{0}}$, yielding the correct
normalization $P(\eta,0;\eta_{0},0)=\delta_{\eta\eta_{0}}$. In the simplest
case of one particle $N=1$, $E_{0}=-1,$ $\langle\eta\rangle_{eff}=\langle
\eta_{0}\rangle_{eff}=1/L$, and we obtain
\begin{equation}
P_{L}(m,t;n,0)=\mathrm{e}^{-t}g_{L}(d,t)=\sum_{\kappa=0}^{\infty}%
\mathrm{e}^{-t}\frac{t^{d+\kappa L}}{(d+\kappa L)!} \label{PoissonPropagator}%
\end{equation}
where $d=m-n$ for $m>n$ and $d=L-n+m$ otherwise, and $n,m$ are initial and
final particle positions. Each term $\mathrm{e}^{-t}t^{d+\kappa L}/(d+\kappa
L)!$ in the sum is a contribution of a Poisson process $\mathrm{e}^{-\lambda
}\lambda^{k}/k!$ where an event (hopping of a particle to the right with rate
$1$) has happened $d+\kappa L$ times during time $t$. Along this trajectory a
particle starts from site $n$, and arrives at site $m$ after making $\kappa$
complete circles on the ring of size $L$. Thus the parameter $\kappa$ in
(\ref{g(d)}) is the winding number.

The formula (\ref{CondProbRes}) has been obtained for the continuous time
Markov process (\ref{MasterEquation}). For the discrete time update
\cite{PriezzhevUpdateBrankov06} we expect the Poisson distribution terms in
(\ref{g(d)}) to be substituted with the binomial distribution terms%
\begin{equation}
g_{L}^{\text{DISCRETE}}(d,t)=\sum_{\kappa=0}^{\infty} \left[  (-1)^{\kappa
}sign(d)\right]  ^{N+1}\binom{t}{d_{L}+\kappa L}p^{d_{L}+\kappa L}
(1-p)^{t-d_{L}-\kappa L}, \label{g(d) Discrete}%
\end{equation}
where $p$ is the probability of a TASEP particle to hop. Unlike the function
(\ref{g(d)}), the sum (\ref{g(d) Discrete}) is truncated for any finite $t$.

\section{Dynamic structure factor}

\label{structure}

The dynamic structure factor in the large current regime of a periodic chain
with $L$ sites is defined as the Fourier transform of the stationary
correlation function $h_{L}(n_{1},t_{1};n_{2},t_{2};\rho) - \rho^{2}=
\langle\hat{n}_{n_{1}}(t_{1})\hat{n}_{n_{2}}(t_{2})\rangle_{eff} - \rho^{2}$,
where $\hat{n}_{k}(t)=\mathrm{e}^{H_{eff}t}n_{k}\mathrm{e}^{-H_{eff}t}$ are
particle number operators and $\rho= N/L$ is the stationary particle density.
Without losing generality, we assume $n_{2}>n_{1}$. Because of the
translational symmetry and time independence of the Hamiltonian $H_{eff}$, the
correlation function $h$ depends only on the differences $n_{2}-n_{1}=n,$
$t_{2}-t_{1}=t$, i.e. $h(n_{1},n_{2},t_{1},t_{2})\equiv h_{L}(n,\rho,t)$. Thus
we can write the real-space representation of the dynamic structure factor as
$S_{L}(n,\rho,t) = h_{L}(n,\rho,t) - \rho^{2}$. Moreover, by particle-hole
symmetry we have that $S_{L}(n,1-\rho,t) = S_{L}(-n,\rho,t)$ and trivially
$S_{L}(n,0,t) = S_{L}(n,1,t)=0$. Hence we can limit our discussion to the
range $0 < \rho\leq1/2$. In order to simplify notation we drop the dependence
on $\rho$ in the structure function and the correlation function and simply
write $S_{L}(n,t)$ and $h_{L}(n,t)$.

The number operator is diagonal and therefore the formula (\ref{Th}) is
applicable. $h_{L}(n,t)$ can then be calculated in similar manner as the
conditional probabilities in the previous section. We shall present only the
final result (the derivation proceeds analogously to the respective quantum
mechanical calculation of $\langle\sigma_{n_{2}}^{z}(t)\sigma_{n_{1}}%
^{z}(0)\rangle$ in \cite{KorepinTMP93}),%
\begin{equation}
S_{L}(n,t) = \frac{1}{L^{2}} \sum_{k=1}^{N}\mathrm{e}^{-i\alpha_{k}%
n+\varepsilon(\alpha_{k})t}\sum_{l=1}^{L}\mathrm{e}^{i\alpha_{l}%
n-\varepsilon(\alpha_{l})t} -\frac{1}{L^{2}}\sum_{k=1}^{N}\mathrm{e}%
^{i\alpha_{k}n-\varepsilon(\alpha_{k})t}\sum_{l=1}^{N}\mathrm{e}^{-i\alpha
_{l}n+\varepsilon(\alpha_{l})t}. \label{h(n,t)}%
\end{equation}
where the $\alpha_{k}$ are of the form (\ref{quasimomenta}) with the ground
state choice $k_{j}=j$, and $\varepsilon(\alpha_{k})$ is given by one of
expressions (\ref{defaultenergy}-\ref{E_XX0}).

Notice that even though the expression (\ref{h(n,t)}) formally looks like the
corresponding quantum formula in imaginary time, its analytical properties and
limits are crucially different. All the sums in (\ref{h(n,t)}) are strictly
real which can be verified straighforwardly, using symmetricity of the ground
state set (\ref{GSmomenta}) of pseudomomenta $\alpha_{k}$. We remark that for
our default choice $\varepsilon(\alpha_{k})=-\mathrm{e}^{-i\alpha_{k}}$, the
summation from 1 through $L$ in Eq.(\ref{h(n,t)}) attains a simple form by
expanding the exponent. We obtain then
\begin{equation}
\frac{1}{L} \sum_{l=1}^{L}\mathrm{e}^{i\alpha_{l}n-\varepsilon(\alpha_{l})t}=
g_{L}(n,t),
\end{equation}
where $g_{L}(n,t)$ is given by (\ref{g(d)}).

From the two-point correlation function (\ref{h(n,t)}) we compute the dynamic
structure factor
\begin{equation}
\hat{S}_{L}(p,t)=\sum_{n=1}^{L}\mathrm{e}^{-2\pi ipn/L}S_{L}(n,t)
\label{DSFdef}%
\end{equation}
with the integer momentum variable $p\in\{1,2,\dots,L\}$.
Obviously $\hat {S}_{L}(p,t) = \hat{S}_{L}(p+nL,t)$ for any
integer $n$ and $\hat{S}_{L}(0,t) = 0$ which allows us to restrict
the subsequent study of the dynamic structure factor to the range
$p\in\{1,2,\dots,L-1\}$. To evaluate (\ref{DSFdef}) in this range
we first observe that the Fourier transformation turns the
exponentials of the summation variables $\alpha_{k},\alpha_{l}$
into the Kronecker-delta $\delta_{p,k-l}$ . Then we write the
second sum in (\ref{h(n,t)}) (which runs up to $N$) as a sum from
1 to $L$ and subtract the part from $N+1$ to $L$. This yields as
an intermediate expression
\begin{align}
\label{DSF1}\hat{S}_{L}(p,t)  &  = \frac{1}{L} \sum_{k=1}^{N} \left[
\mathrm{e}^{(\varepsilon(\alpha_{k})-\varepsilon(\alpha_{k+p}))t} -
\mathrm{e}^{-(\varepsilon(\alpha_{k})-\varepsilon(\alpha_{k-p}))t}\right]
\nonumber\\
&  + \frac{1}{L} \sum_{k=1}^{N} \sum_{l=N+1}^{L} \mathrm{e}^{-(\varepsilon
(\alpha_{k})-\varepsilon(\alpha_{l}))t} \delta_{p,k-l}\\
&  := \hat{S}_{L}^{(1)}(p,t) + \hat{S}_{L}^{(2)}(p,t)
\end{align}
for which we analyse next the double sum $\hat{S}_{L}^{(2)}(p,t)$. We focus on
the default case (\ref{defaultenergy}).

For the default case the difference of relaxation times in the exponential
takes the simple form
\begin{equation}
\varepsilon(\alpha_{k})- \varepsilon({\alpha_{l}}) = - (1-\mathrm{e}^{2\pi i
p/L})\varepsilon(\alpha_{k})
\end{equation}
and for notational convenience we introduce
\begin{equation}
t_{p} := (1-\mathrm{e}^{2\pi i p/L}) t.
\end{equation}
Bearing in mind the range of definition of the momentum variable
$p\in\{1,2,\dots,L-1\}$, the Kronecker-delta in conjunction with
the summation limits of the double sum gives rise to three
distinct regimes for $p$. Careful analysis yields
\begin{equation}
\label{DSF2}\hat{S}_{L}^{(2)}(p,t) = \left\{
\begin{array}
[c]{ll}%
\displaystyle \frac{1}{L} \sum_{k=1}^{p} \mathrm{e}^{t_{p}\mathrm{e}%
^{-i\alpha_{k}}} & p = 1,\dots, N-1\\
\displaystyle \frac{1}{L} \sum_{k=1}^{N} \mathrm{e}^{t_{p}\mathrm{e}%
^{-i\alpha_{k}}} & p = N,\dots, L-N\\
\displaystyle \frac{1}{L} \sum_{k=N+1-L+p}^{N} \mathrm{e}^{t_{p}%
\mathrm{e}^{-i\alpha_{k}}} & p = L-N+1,\dots, L-1.
\end{array}
\right.
\end{equation}

\begin{figure}[ptb]
\centerline{ \subfigure[\label{figEven}]{
\includegraphics[width=7.5cm,height=5.3cm,clip]{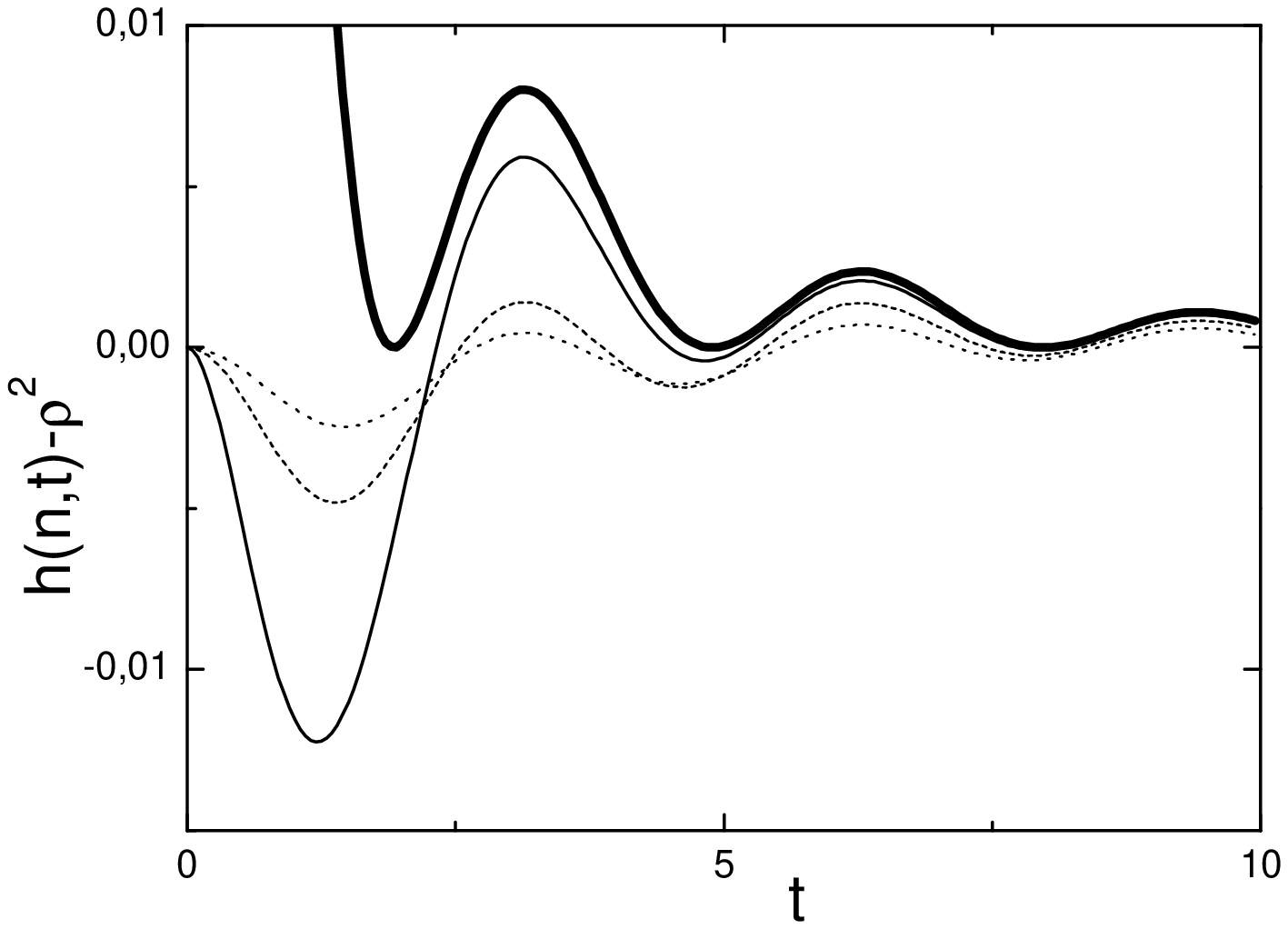}}
\subfigure[\label{figOdd}]{
\includegraphics[width=7.6cm,height=5.2cm,clip]{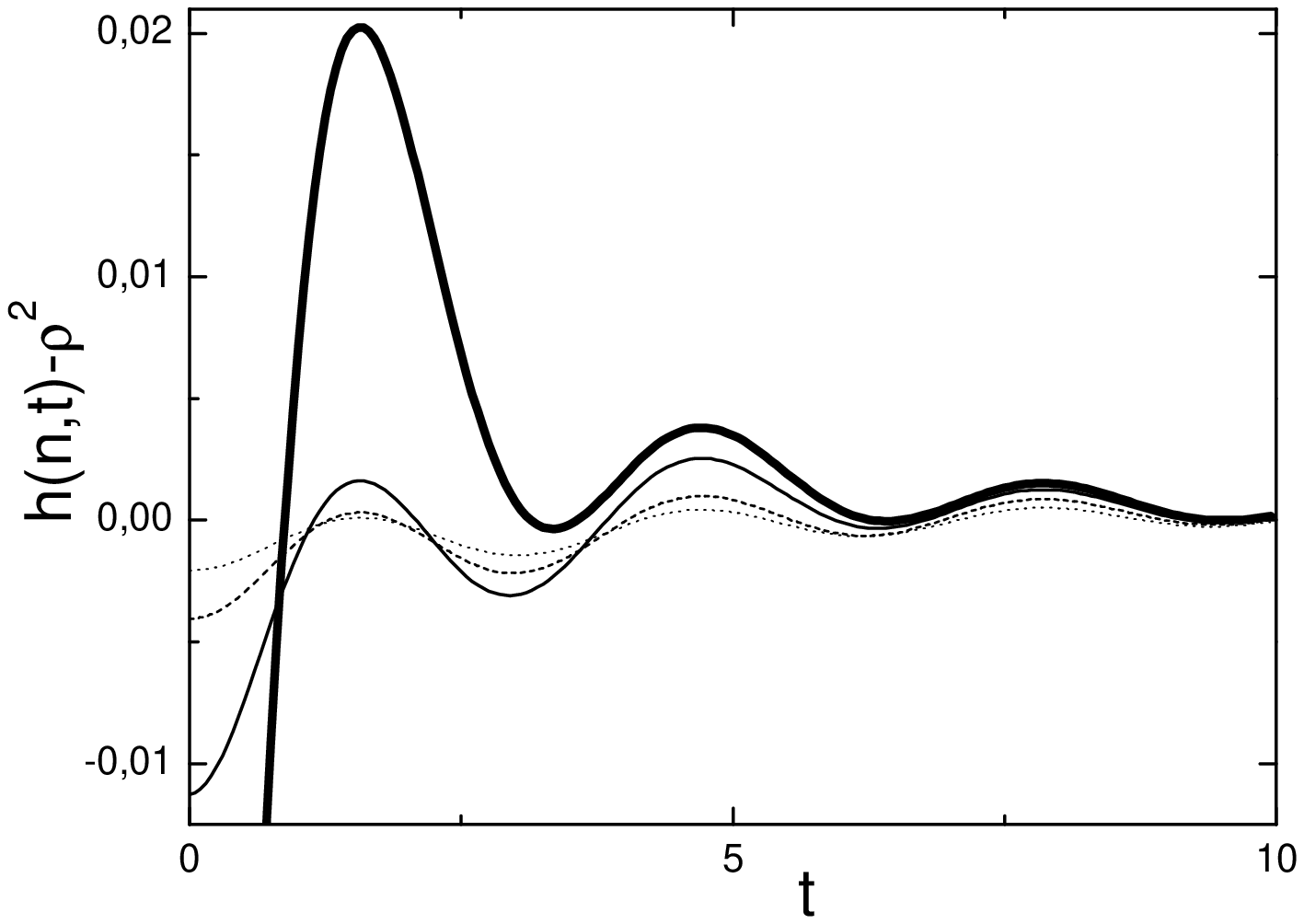}}
} \caption{Dynamic structure factors $\langle n_{m}(t) n_{0}(0) \rangle-
\rho^{2}$ for $\rho=0.5$ as function of time for $0 \leq n < 8 $, computed
from Eq.(\ref{h(n,t)ThermodynamicLimit}). \textbf{Panel (a)}: connected
correlation function $h(n,t)-\rho^{2}$ for odd $n=0,2,4,6$, represented by
bold, thin, dashed, and dotted lines respectively. \textbf{Panel (b)}:
connected correlation function $h(n,t)-\rho^{2}$ for odd $n=1,3,5,7$( bold,
thin, dashed, and dotted lines respectively). }%
\label{Fig_EvenOdd}%
\end{figure}

In the thermodynamic limit $L\gg1$ the sums over $k$ and $j$ in (\ref{h(n,t)})
turn into integrals, $R(n,\rho,t):=\lim_{L\to\infty} \frac{1}{L}\sum_{k=1}%
^{N}\mathrm{e}^{i\alpha_{k} n-\varepsilon(\alpha_{k})t}=\frac{1}{2\pi}%
\int_{-\pi\rho}^{\pi\rho}\mathrm{e}^{ipn-\varepsilon(p)t}dp$. Notice that here
$p$ is real-valued and we define it to be in the interval $[-\pi,\pi]$. This
yields
\begin{equation}
S(n,t) := \lim_{L\to\infty} S_{L}(n,t) = R(n,1,t)R(-n,\rho,-t)-R(n,\rho,t)R(-n,\rho,-t).
\label{h(n,t)ThermodynamicLimit}%
\end{equation}
where $R(n,1,t)=t^{n}/n!$ corresponds to the limit of the function $g(n,t)$,
where only the first term in (\ref{g(d)}) appears (no winding condition). With
a view in large, but still finite systems we point out that this observation
imposes obvious validity limitations on the integral expression
(\ref{h(n,t)ThermodynamicLimit}) as an approximation for large system size:
The limiting behaviour cannot be used as an approximation for $t\gtrsim L$ and
$n\gtrsim L$. The integrals $R(n,\rho,t)$, apart from obvious special cases
$R(n,1,t)=t^{n}/n!$, and $R(n,\rho,0)=(n\pi)^{-1}\sin n\pi\rho$ are not
expressed in elementary functions and must be evaluated numerically. In
Fig.~\ref{Fig_EvenOdd} we show the function $S(n,t)$ for even and odd $n$ for
the particular case of half-filling $N/L=1/2$ at different times $t$. For
$t=0$ the difference in the expression $\varepsilon_{q}$ for energies the
quasiparticles become irrelevant and, using $R(n,\rho,0)=(n\pi)^{-1}\sin
n\pi\rho$, we obtain from (\ref{h(n,t)ThermodynamicLimit}) the static
density-density correlation function
\begin{equation}
\label{staticcorrelation}
S(n,0) = -\frac{\sin^{2}n\pi\rho}{n^{2}\pi^{2}}%
\end{equation}
first derived in \cite{LiebSchultzMattis61}.

In order to explore the large-scale behaviour of the dynamic structure factor
we study the behaviour for small momentum $p$ and large times $t$. To this end
we return to the definition (\ref{DSFdef}) and first observe that in the
thermodynamic limit
\begin{equation}
\label{DSF3}\hat{S}^{(1)}(p,t) = \frac{1}{2\pi} \int_{-\rho\pi}^{\rho\pi}
dx\,\left[  \mathrm{e}^{-\mathrm{e}^{-ix}t_{-p}} - \mathrm{e}^{\mathrm{e}%
^{-ix}t_{p}} \right]
\end{equation}
where now $t_{p} = (1-\mathrm{e}^{ip})t$. Using $t_{-p} = - \mathrm{e}^{-ip}
t_{p}$ allows us to rewrite this expression as a difference of integrals over
the same function $\exp({\mathrm{e}^{-ix}t_{p})}$ with integration intervals
$[-\rho\pi+p,\rho\pi+p]$ (positive term) and $[-\rho\pi,\rho\pi]$ (negative
term) respectively. On the other hand
\begin{equation}
\label{DSF4}\hat{S}^{(2)}(p,t) = \left\{
\begin{array}
[c]{ll}%
\displaystyle \frac{1}{2\pi} \int_{-\rho\pi}^{-\rho\pi+p} dx\, \mathrm{e}%
^{t_{p}\mathrm{e}^{-ix}} & p \in[0,2\rho\pi]\\
\displaystyle \frac{1}{2\pi} \int_{-\rho\pi}^{\rho\pi} dx\, \mathrm{e}%
^{t_{p}\mathrm{e}^{-ix}} & p \in[-\pi,\dots,-2\rho\pi] \cup[2\rho\pi,\dots
,\pi]\\
\displaystyle \frac{1}{2\pi} \int_{\rho\pi+p}^{\rho\pi} dx\, \mathrm{e}%
^{t_{p}\mathrm{e}^{-ix}} & p \in[-2\rho\pi,0]
\end{array}
\right.
\end{equation}
Putting everything together we finally obtain the dynamic structure factor for
$\rho\leq1/2$
\begin{equation}
\label{DSFexact}\hat{S}(p,t) = \left\{
\begin{array}
[c]{ll}%
\displaystyle \frac{1}{2\pi} \int_{\rho\pi}^{\rho\pi+p} dx\, \mathrm{e}%
^{t_{p}\mathrm{e}^{-ix}} & p \in[0,2\rho\pi]\\
\displaystyle \frac{1}{2\pi} \int_{-\rho\pi+p}^{\rho\pi+p} dx\, \mathrm{e}%
^{t_{p}\mathrm{e}^{-ix}} & p \in[-\pi,\dots,-2\rho\pi] \cup[2\rho\pi,\dots
,\pi]\\
\displaystyle \frac{1}{2\pi} \int_{-\rho\pi+p}^{-\rho\pi} dx\, \mathrm{e}%
^{t_{p}\mathrm{e}^{-ix}} & p \in[-2\rho\pi,0]
\end{array}
\right.
\end{equation}
This, along with the symmetry relation $\hat{S}(1-\rho,p,t) = \hat{S}%
(\rho,-p,t)$ provides an exact integral presentation valid for all densities
$\rho$, momenta $p$ and times $t$. The static structure factor takes the
simple form ($\rho\leq1/2$)
\begin{equation}
\label{DSFstatic}\hat{S}(p,0) = \left\{
\begin{array}
[c]{ll}%
\displaystyle \frac{|p|}{2\pi} & p \in[-2\rho\pi,2\rho\pi]\\
\displaystyle \rho & p \in[-\pi,\dots,-2\rho\pi] \cup[2\rho\pi,\dots,\pi],
\end{array}
\right.
\end{equation}
cf. the real-space result (\ref{staticcorrelation}).

We are particularly interested in the large scale behaviour as
expressed in the scaling limit of small $p$ and large $t$ of the
form $p^{z}t = u$ where $z$ is the dynamical exponent and $u$ is
the scaling variable. In the limit $p\to0$ only the first and
third expression in (\ref{DSFexact}) are relevant. From the
occurrence of the factor $t_{p}=(1-\mathrm{e}^{ip})t$ we conclude
that there is non-trivial scaling behaviour for $z=1$, i.e. for
$u=pt$. In this scaling we have $t_{p} = -iut$ which yields the
desired result
\begin{equation}
\label{DSFscaling}\hat{S}(u) = \frac{|u|}{2\pi t} \mathrm{e}^{-iu \cos{\rho
\pi} - |u|\sin{\rho\pi}}%
\end{equation}
which is valid for all $\rho\in[0,1]$  and in agreement with the
universal form the dynamic structure factor derived in
\cite{Spoh99} for the symmetric case. The presence of the particle
drift does not change the universality class as it does for the
usual unconditioned exclusion process where the undriven model is
in the universality class of the Edwards-Wilkinson equation with
dynamical exponent $z=2$ while the driven model is in the KPZ
universality class with $z=3/2$. This is in agreement with an
earlier observation that for stochastic dynamics which have an
underlying free-fermion structure an external drift can be
absorbed into a Galilei transformation \cite{Schu95}.


\section{Final remarks}

We obtained analytically conditional probabilities and the two
point time-dependent density correlation functions for the ASEP
conditioned to carry a very large average current. The conditional
probabilities have determinantal form and can be expressed through
elementary functions. The density correlation functions are
obtained both for a finite system and in the thermodynamic limit.
By Fourier transformation we have computed the exact dynamical
structure factor and derived its large-scale behaviour. The
natural scaling variable turns out to be $u=pt$ which proves that
the dynamical exponent of the conditioned ASEP is $z=1$. From the
explicit scaling form we read off the collective velocity $v_{c} =
\cos{\rho\pi}$ of density fluctuations which is in contrast to
$v_{c} = (p-q)(1-2\rho)$ of the usual ASEP in the regime of
typical currents. The relaxation part is symmetric in $u$ and very
different from the corresponding quantity in the usual ASEP
\cite{Prae02} which has dynamical exponent $z=3/2$ for the driven
case and $z=2$ for the symmetric case.  In our model the presence
of a drift does not change the universality class.

Our results have a natural generalization to the study of large activity, i.e.
to choosing the Hamiltonians (\ref{ASEPactivity}) and (\ref{XX0}). One has to
replace the energies $\varepsilon(\alpha_{k})$ in (\ref{h(n,t)}) by the
respective expressions $-\left(  p\mathrm{e}^{-i\alpha_{k}}+r\mathrm{e}%
^{i\alpha_{k}}\right)  $ and $-2\cos\alpha_{k}$.  It will be
interesting to study not only the hydrodynamic limit, but also the
microscopic structure of shocks for the general case. It would
also be interesting to extend the analysis of the effective
dynamics of driven systems under large deviation constraints to
the non-diagonal case, e.g. to compute current-current
time-dependent averages.

\section*{Acknowledgements}

We thank M. Salerno and H. Spohn for stimulating discussions and D. Simon for valuable
comments on a preliminary version of the manuscript. This work was supported by
Deutsche Forschungsgemeinschaft.


\end{document}